\documentclass[%
 reprint,
 amsmath,amssymb,
 aps,
]{revtex4-1}

\usepackage{graphicx}
\usepackage{dcolumn}
\usepackage{bm}

\begin{document}

\preprint{APS/123-QED}

\title{Splash Detail Due to Grain Incident on Granular Bed}

\author{Takahiro Tanabe$^1$}
\author{Tomoki Koike$^2$}%
\author{Takashi Shimada$^2$}%
\author{Nobuyasu Ito$^2$}%
\author{Hiraku Nishimori$^1$}
 \email{nishimor@hiroshima-u.ac.jp}
\affiliation{%
 $^1$Department of Mathematical and Life Science, Graduate school of Science, Hiroshima University, Higashihiroshima, Hiroshima 739-8526, Japan\\
 $^2$Department of Applied Physics, Graduate school of Engineering, The University of Tokyo 7-3-1 Hongo, Bunkyo, Tokyo 113-0033, Japan
}%




\date{\today}

\begin{abstract}
{
Using the discrete element method (DEM), we study the splash processes induced by the impact of a grain on two types of granular beds, namely, randomly packed and FCC-structured beds.
Good correspondence is obtained between our numerical results and the findings of previous experiments, and it is demonstrated that the packing structure of the granular bed strongly affects the splash process.
The mean ejection angle for the randomly packed bed is consistent with previous experimental results. 
The FCC-structured bed yields a larger mean ejection angle; however, the latter result has not been confirmed experimentally.
Furthermore, the ejection angle distributions and the vertical ejection speeds for individual grains vary depending on the relative timing at which the grains are ejected after the initial impact.
Obvious differences are observed between the distributions of grains ejected during the earlier and later splash periods: the form of the vertical ejection speed distribution varies from a power-law form to a lognormal form with time, and more than 80\% of the kinetic energy of all ejected grains is used for earlier ejected grains.
}
\end{abstract}

\pacs{Valid PACS appear here}
\maketitle
\section{Introduction}
Massive sediment transport phenomena, such as dust storms and drifting snow, pose a considerable threat to human life. Further, the formation of geomorphological patterns on sand-desert and snowfield surfaces as a result of sediment transport, such as dunes and ripples, is of considerable research interest.
To elucidate the granular transport that occurs near the surfaces of sand deserts and snow fields, it is necessary to focus on the collisions between wind-blown grains and these surfaces along with the resultant ejection of grains from the surfaces.
This approach is merited because, in the case of wind-blown grain transport, the major component of the grain entrainment into the air is caused by both the collision and ejection¥cite{Bagnold,Sus}.
This mechanism is called the ``splash process.''

Splash processes have been widely studied using various techniques.
For example, Werner {\it et al.} have simulated grain-bed collision processes in a two-dimensional system\cite{Werner}, while 
Nishida {\it et al.} have performed numerical simulations of granular splash behavior in a three-dimensional (3D) system and analyzed the relation between the impact and ejection angles ($\theta_I$ and $\theta_E$, respectively) projected onto the surface of a granular bed\cite{Nishida1}. 
Further, Xing and He have performed 3D collision simulations with mixed binary grains\cite{Mao}, and Wada {\it et al.} have numerically modeled the impact cratering process on a granular target\cite{Wada}.
In a physical experiment, Katsuragi {\it et al.} created small-scale craters in a laboratory system\cite{Katsuragi}, whereas
Sugiura {\it et al.} estimated the splash function of snow grains via wind-tunnel experiments\cite{Sugiura1,Sugiura2}.
In addition, Ammi {\it et al.} performed a 3D splash experiment and recorded the results using two high-speed cameras, demonstrating that the mean ejection angle $\overline{\theta}_E$ of a series of splashed grains is independent of both $\theta_I$ and the velocity of the incident grains $\mbox{\boldmath$V$}_I$, and it is close to $60^\circ$ \cite{Ammi}.
In their experiment, a randomly packed (RP) bed was considered, and the final result suggests that the behavior at the first instance of impact during a splash process involving a granular bed has no influence on the later behavior.

In the present study, we perform numerical simulations in order to investigate the splash processes in more detail.
Assuming that the packing structure of a granular bed affects the splash behavior, we consider not only an RP bed (an RP bed corresponds to the scenario examined in the experiment of Ammi {\it et al.}\cite{Ammi}, except for differences in the dimensions of the simulation space and the grain features), but also an FCC-structured bed (hereafter, ``FCC bed").
Thus, we analyze the dependence of the splash process on the bed structure.
In addition, we investigate the details of the ejection grains for each splash paying attention to their ejection timing.

\section{Model}
\begin{figure}[tb]
\begin{center}
\includegraphics[height=3.0cm]{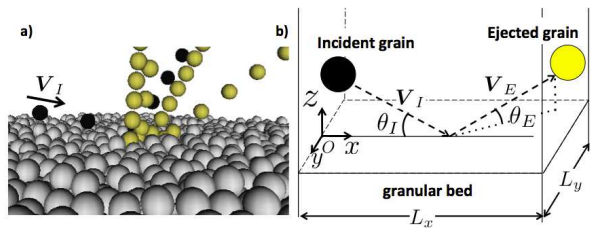}
\caption{a) Snapshot of the splash process. The black and yellow chains represent the motion of the incident and ejected grains, respectively.
b) Definitions of the incident angle $\theta_I$, incident velocity $\mbox{\boldmath$V$}_I$, ejection angle $\theta_E$, and ejection velocity $\mbox{\boldmath$V$}_E$.}
\label{f1}
\end{center}
\end{figure}
\subsection{Basic Setup}
In this study, the splash processes are examined using the discrete element method (DEM).
The translational movement of the grains obeys Newton's law of motion, and grain rotation is neglected. 
Thus, the equation of motion is
\begin{equation}
\label{eq:Newton}
m_i\ddot{\mbox{\boldmath$x$}}_i=\sum_j(\mbox{\boldmath$F$}_H^{i,j}+\mbox{\boldmath$F$}_D^{i,j})-m_ig\mbox{\boldmath$e$}_z,
\end{equation}
where $\mbox{\boldmath$x$}_i$ and $m_i$ are the position and the mass of the {\it i}-th grain, respectively; and $g$ and $\mbox{\boldmath$e$}_z$ are the gravity constant and the vertical unit vector in the upward direction, respectively.
$\mbox{\boldmath$F$}_H^{i,j}$ and $\mbox{\boldmath$F$}_D^{i,j}$ represent the repulsive and dissipative forces acting between the {\it i}-th and {\it j}-th grains, respectively, as explained in Sect. 2.2.
Our ``simulation box'' consists of a fixed bottom and walls, which make up a roofless 3D cubic container (Fig.\ref{f1}).
The walls and the bottom floor are made of the same material as the grains.
Two types of initial granular bed structures are prepared (RP and FCC), as explained in greater detail in Sect. 2.3.

A grain is fired at the bed at a certain incident angle $\theta_I$ and incident speed $V_I$ (Fig. \ref{f1}(b)).
As a result of the collision between the projectile grain and the granular bed, a number of grains are expelled from the bed.
Here, we define the initially projected grain as the ``incident grain'' and the expelled grains that reach a certain threshold height (see Sect. 2.4) as the ``ejected grains'' (Fig. \ref{f1}(a)).
We exclude the rebounding incident grain from consideration as an ejected grain.
In this study, we consider monodispersed grains only; therefore, all of the grains comprising the granular bed and the incident grain have the same mass and radius.
The parameters used in the simulation are summarized in Table \ref{t1}.
\begin{table}[b]
\caption{Simulation parameters}
\label{t1}
\begin{center}
\begin{tabular}{ll}
\hline
\multicolumn{1}{c}{Parameter} & \multicolumn{1}{c}{Value} \\
\hline
\verb|System size (bottom area)| & $35\times35\,{\rm cm^2}$ \\
\verb|Gravity| & $10\,{\rm m/s^2}$ \\
\verb|Young's modulus| & $1.0\times10^9\,{\rm kg/ms^2}$ \\
\verb|Grain radius| & 0.5 {\rm cm}\\
\verb|Grain mass| & 0.1 {\rm g}\\
\hline
\end{tabular}
\end{center}
\end{table}
\subsection{Grain Interaction}
We treat grains as viscoelastic spheres.
For the elastic force, we adopt the Hertzian force $\mbox{\boldmath$F$}^{i,j}_H$\cite{Hertz}, with
\begin{equation}
\mbox{\boldmath$F$}_H^{i,j}=-k_n\sqrt{\frac{r_ir_j}{r_i+r_j}\delta_n^{i,j}}\hspace{1mm}\mbox{\boldmath$\delta$}_n^{i,j},
\label{eq:hertzian}
\end{equation}
where $k_n,r_i$, and $\delta_n^{i,j}=|\mbox{\boldmath$\delta$}_n^{i,j}|$ are the Young's modulus, the radius of the {\it i}-th grain, and the displacement from the natural contact position $r_i+r_j$, respectively. Further,
\begin{equation}
\mbox{\boldmath$\delta$}_n^{i,j}=
\begin{cases}
\mbox{\boldmath$0$,} & {\rm(non\hspace{-1mm}-\hspace{-1mm}contact)},\\
\Big[|\mbox{\boldmath$x$}_i-\mbox{\boldmath$x$}_j|-(r_i+r_j)\Big]\mbox{\boldmath$n$}_{i,j}, &{\rm{(contact)}},
\end{cases}
\label{eq:displacement}
\end{equation}
where $\mbox{\boldmath$n$}_{i,j}=(\mbox{\boldmath$x$}_i-\mbox{\boldmath$x$}_j)/|\mbox{\boldmath$x$}_i-\mbox{\boldmath$x$}_j|$ is the unit vector in the normal direction.

To represent the energy dissipation, we adopt the friction force $\mbox{\boldmath$F$}^{i,j}_D$, with 
\begin{equation}
\mbox{\boldmath$F$}_D^{i,j}=-\eta_n\mbox{\boldmath$v$}_n^{i,j}=\\
-\alpha\sqrt{m^*k_n\sqrt{\frac{r_ir_j}{r_i+r_j}\delta_n^{i,j}}}\hspace{1mm}\mbox{\boldmath$v$}_n^{i,j},
\label{eq:disspation}
\end{equation}
where $\mbox{\boldmath$v$}_n^{i,j}$, $\eta_n$, and $m^*$, are the relative normal velocity, damping coefficient, and the reduced mass, respectively. Note that $\alpha$ is relative to the restitution coefficient $e$ \cite{Tuji}.
In our simulation, the value of $e$ is fixed at 0.9.

When the grains reach the boundaries, the sum of the Hertzian and friction forces acts on the grains such that the work force $\mbox{\boldmath$F$}_W^i$ is expressed as
\begin{equation}
\mbox{\boldmath$F$}_W^i=-k_n\sqrt{r_i\delta_n^i}\mbox{\boldmath$\delta$}_n^i-\eta_{n,w}\mbox{\boldmath$v$}_n^i,
\label{eq:boundary}
\end{equation}
where $\eta_{n,w}$ is the damping coefficient between grains and boundaries.
This equation corresponds to Eqs. (\ref{eq:hertzian}) and (\ref{eq:disspation}), with the plain wall limit: $r_j\to\infty$.
\subsection{Packing Structure}
We construct the initial RP and FCC beds as follows.
The RP bed is created through the free falling of 32,768 grains. At first, all grains are placed at random positions in the simulation box, with no overlap.
Then, they fall to the bottom as a result of the effects of $g$, losing kinetic energy through the dissipative repulsive force of Eqs. (\ref{eq:hertzian})--(\ref{eq:boundary}).
The packing process is completed after a sufficient relaxation time has elapsed.

On the other hand, the initial positions of the grains in the FCC bed are approximately determined, except for the fine tuning of their positions according to $g$ and the nonlinear interactions of Eqs. (\ref{eq:hertzian})--(\ref{eq:boundary}).
Similar to the previous RP procedure, the packing of the FCC structure is completed after a sufficient relaxation time has elapsed.
The volume fractions of the RP and FCC beds are approximately 0.63 and 0.74, respectively.
In previous experiments with monodispersed spherical beads, the volume fraction of the grains was approximately 0.6 in RP beds\cite{Wada,Ammi,Rioul}.
It has been reported for a two-dimensional system that a bed thickness of more than 24 layers is needed to exclude the shockwave effects.
The average height of the RP bed surface is approximately 22 grains.
To construct the FCC bed, 36,639 grains and a 24-layer pile are used.

\subsection{Definition of Injection and Ejection}
The pair of $V_I$ and $\theta_I$ characterize the injection of the incident grain (Fig. \ref{f1}).
For a given $V_I$ and $\theta_I$, the incident velocity is determined from $\mbox{\boldmath$V$}_I=(V_I\cos\theta_I,0,-V_I\sin\theta_I)$.
In this study, $V_I$ is set to 10.0, 25.0, or 40.0 m/s while $\theta_I$ is varied among $10^\circ, 40^\circ, 60^\circ$, and $90^\circ$.
To obtain sufficient data for statistically meaningful results, 100 splash simulations are conducted for each set of $(V_I, \theta_I)$, with different initial positions.

The horizontal coordinate of the ``collision point'' $(x_c,y_c,z_c)$ between the incident grain and the granular bed surface is given randomly within a central horizontal circle on the bed surface, which we call the ``incident circle.''
The radius of this circle is three times the grain diameter.
The center of the incident circle is $(L_x/2,L_y/2,z_b)$, where $L_x$ and $L_y$ are the lengths of the $x$ and $y$ sides of the simulation box, respectively (Fig. \ref{f1}(b)), and $z_b$ is the highest $z$ coordinate of the grain surface (the upper edge of the highest grain) within the above-mentioned incident circle.
The initial position of the incident grain (grain center) is $(x_0,y_0,z_0)=(x_c-t_cV_I\cos\theta_I, y_c, z_b+2r)$, where $r$ is the radius of the incident grain, and $t_c=\left(-V_I\sin\theta+\sqrt{V^2_I\sin\theta^2+4gr}\right)\Big{/}g$ is the time required for collision with the surface, calculated from the given $\mbox{\boldmath$V$}_I$.
We define grains with centers that reach $z_{\rm th}$ as ``ejected grains'' and record their ejection velocity $\mbox{\boldmath$V$}_E$, where $z_{\rm th}$ is $2r$ above the average bed surface height around the contact point.
Furthermore, we define the rebound of the incident grain (``rebound grain'') and its velocity $\mbox{\boldmath$V$}_R$ with same definition as the ejected grains.
This ejected-grain criterion roughly corresponds to those of previous 3D splash experiments\cite{Ammi,Rioul}.
In this paper, we also define $\theta_E$ as
\begin{equation}
\theta_E=\arccos\left(
\frac{\sqrt{V^2_{E,x}+V^2_{E,y}}}{\sqrt{V^2_{E,x}+V^2_{E,y}+V^2_{E,z}}}
\right),
\end{equation}
where $V_{E,x},V_{E,y}$, and $V_{E,z}$ are the components of $\mbox{\boldmath$V$}_E$ (Fig. \ref{f1}(b)).

\begin{figure}[t]
\begin{center}
\includegraphics[scale=0.13]{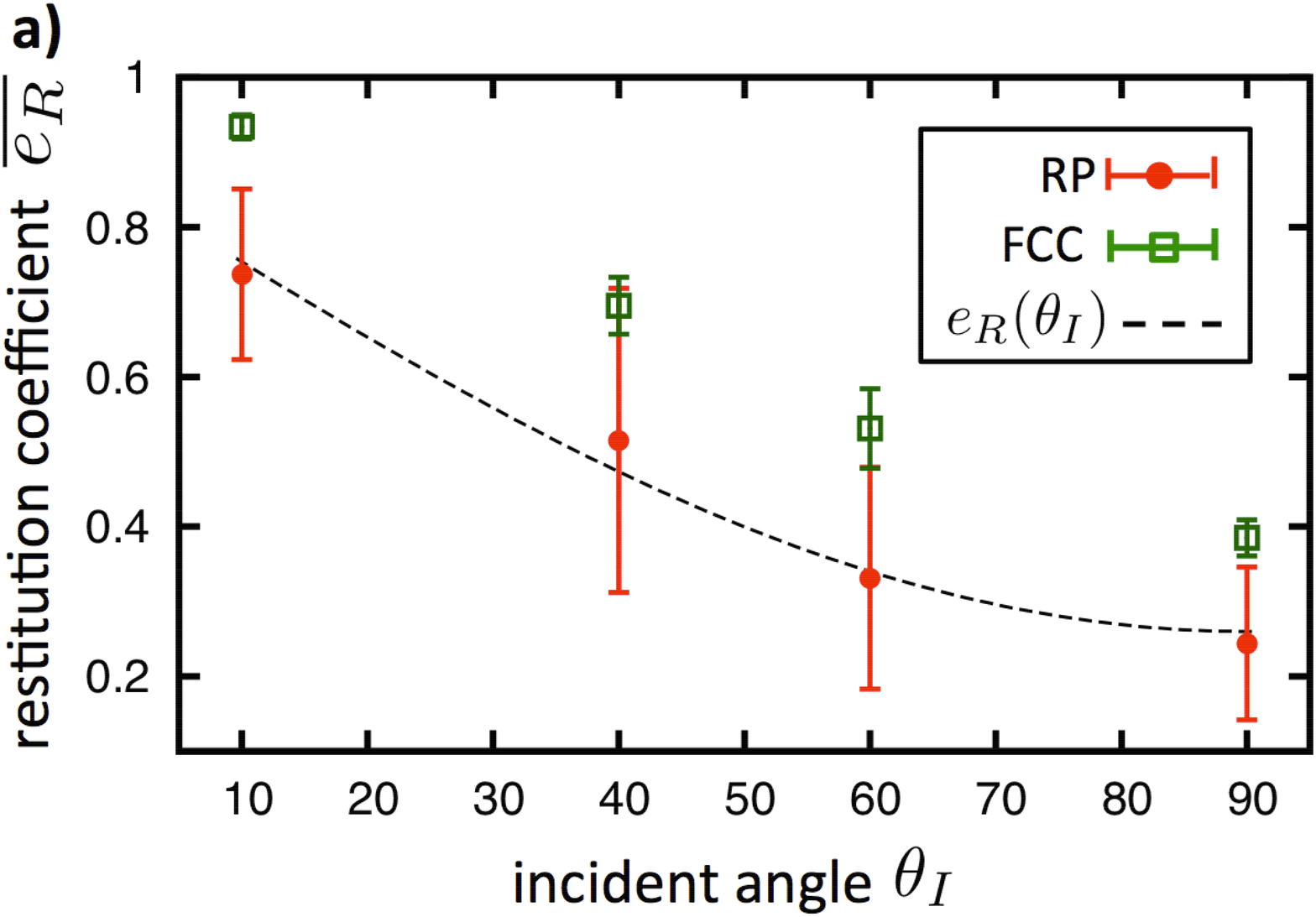}
\includegraphics[scale=0.13]{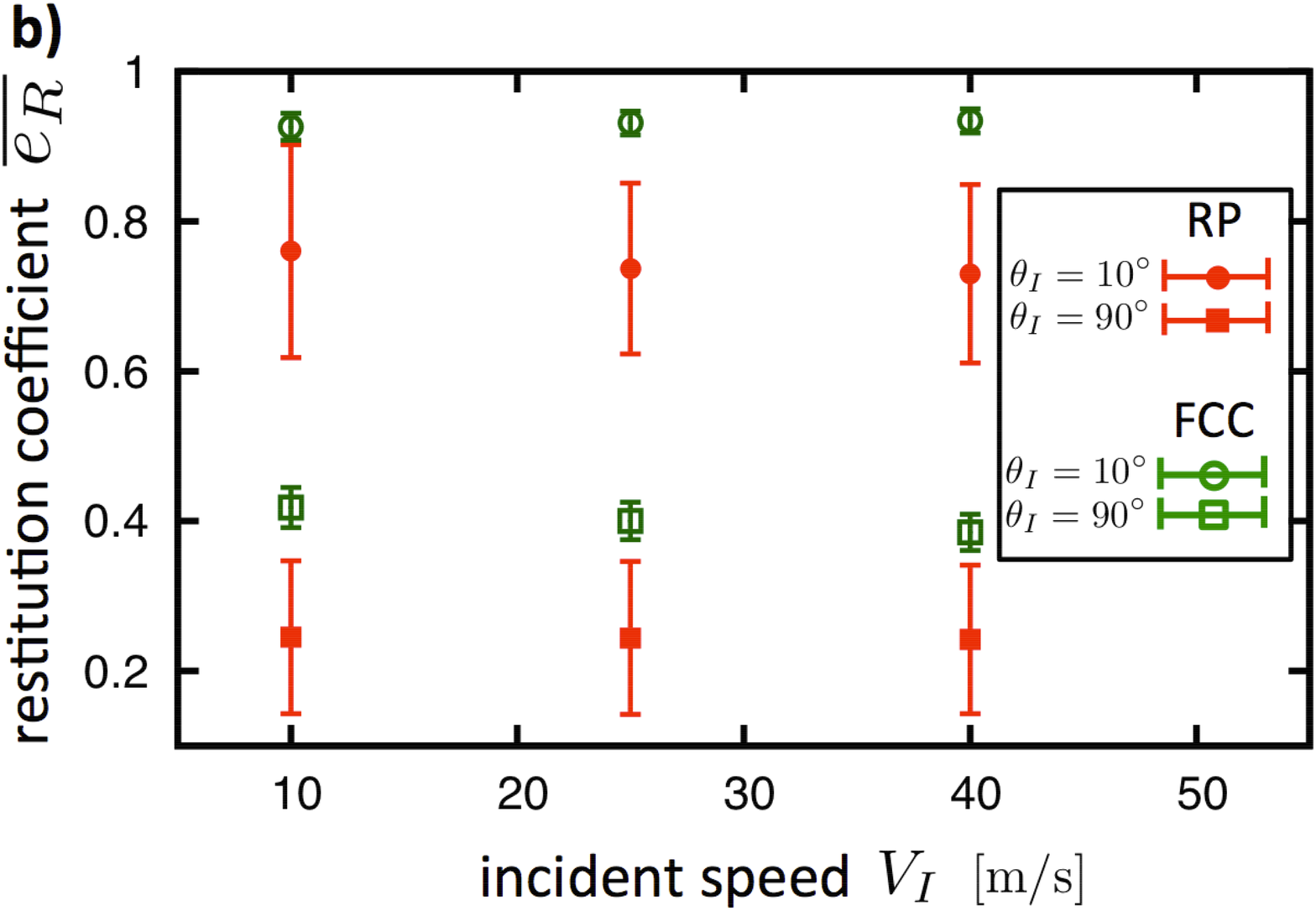}
\end{center}
\caption{Mean restitution coefficient of the incident grain $\overline{e_R}=\overline{V_R}/V_I$ for (a) various incident angles $\theta_I$ ($V_I=25.0{\rm m/s}$: fixed) and (b) incident speeds $V_I$ ($\theta_I=10^\circ$ (circle) and $90^\circ$ (square): fixed) in RP (filled symbol) and FCC (open symbol) beds.
The error bars indicate standard deviations, and the dashed line is the best fit of the form $A-R\sin\theta_I$ for the RP bed ($A\approx0.86$ and $B\approx0.60$).}
\label{Energy}
\end{figure}
\section{Results}
\subsection{Incident Energy}
The mean incident energy $\overline{E_b}$, which means the energy transferred from the incident grain to the granular beds, is important to consider for the ejected grains.
Since $\overline{E_b}$ is equal to the energy lost by the incident grain, we obtain the following relation: $\overline{E_b}=E_I-m\overline{V_R}^2/2=E_I(1-\overline{e_R}^2)$, where 
$E_I=mV_I^2/2$, $\overline{e_R}=\overline{V_R}/V_I$, and $\overline{V_R}$ is the mean speed of the incident grain at $z=z_{\rm th}$ after impact.
Therefore, we focus on $\overline{e_R}$ to characterize the incident energy transferred to the bed.
Figure \ref{Energy} shows that $\overline{e_R}$ only depends on the incident angle and does not depend on the incident speed; these results reproduce those in previous experiments\cite{Werner,Ammi}.
Our result obtained for the RP bed corresponds well with the fitting function 
\begin{equation}
e_R(\theta_I)=A-B\sin\theta_I
\end{equation}
which was proposed in a previous study\cite{Ammi} (Fig. \ref{Energy}(a)).
In our simulation, $A\approx0.86$ and $B\approx0.60$ for $V_I=25.0 {\rm m/s}$; these values are close to those from previous collision experiments.
The value of $\overline{e_R}$ obtained for the FCC bed is larger than that for the RP bed for same pair of $V_I$ and $\theta_I$.
Because of the lower roughness of the FCC bed surface, the error bars are very small.
From aforementioned results, the fraction of incident energy $(1-\overline{e_R}^2)$ increases with $\theta_I$ and is independent of $V_I$, and $(1-\overline{e_R}^2)$ for the FCC bed is smaller than that for the RP bed.

\begin{figure}[t]
\begin{center}
\includegraphics[height=4.5cm]{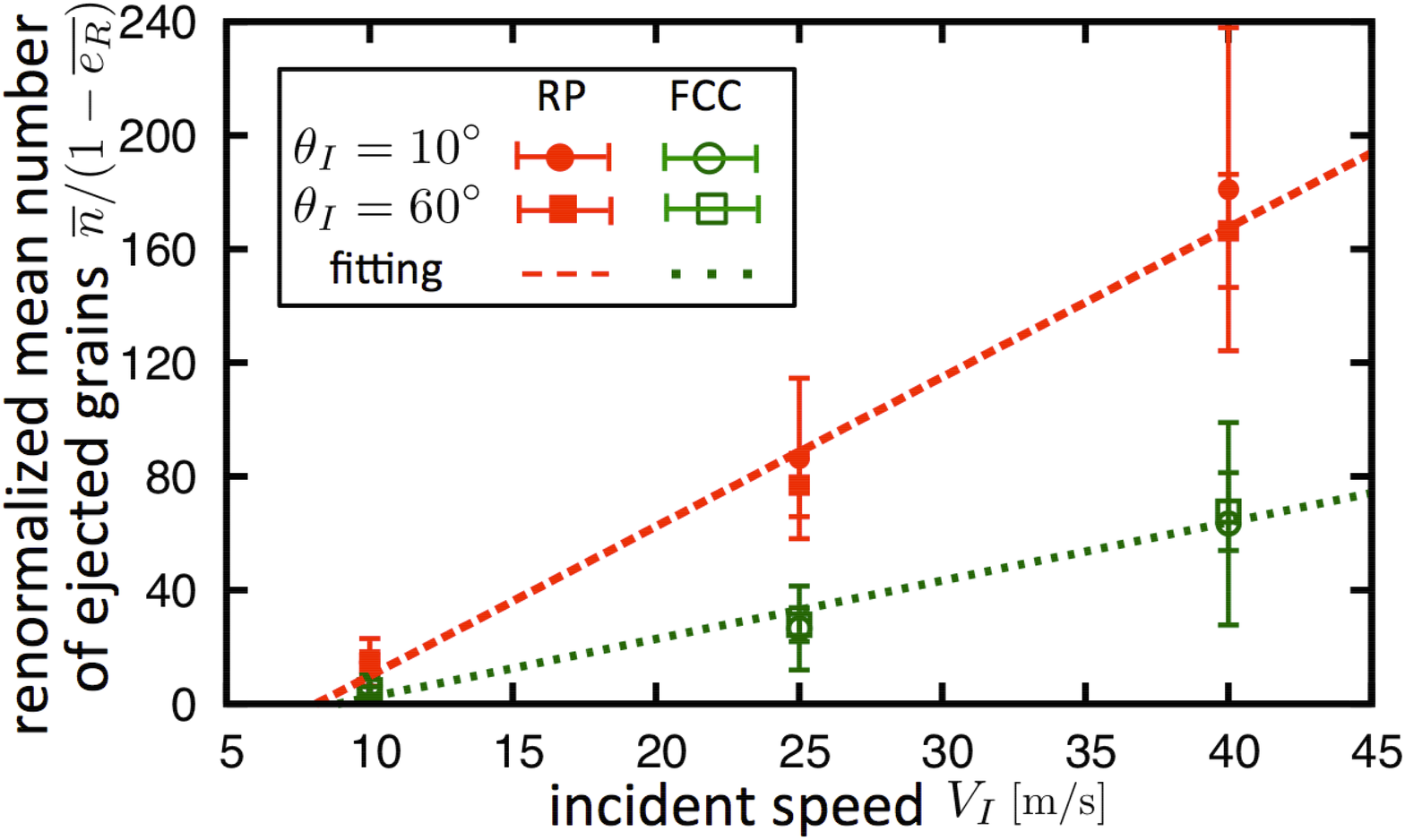}
\caption{
Mean number of ejected grains renormalized by $(1-\overline{e_R}^2)$ ($\overline{e_R}$ is the mean restitution coefficient of the rebounded incident grain) versus the incident speed $V_I$ for various incident angles $\theta_I$ (circles: $10^\circ$, and squares: $60^\circ$).
The filled and open symbols correspond to RP and FCC beds, respectively.
The dashed lines are fits based on Eq. (\ref{eq:num}).
The error bars indicate standard deviations.
}
\label{num}
\end{center}
\end{figure}
\subsection{Ejection Number}
The number of ejected grains after each splash process is related to the amount of kinetic energy transferred from the incident grain to the granular bed.
Kinetic energy propagates into the granular bed, in which the energy is dissipated via the interactions between the grains.
Because increases in $\theta_I$ and $V_I$ produce a high value of $\overline{E_b}$, the ensemble averages of the mean number of ejected grains for each splash $\overline{n}$ increase with $\theta_I$ and $V_I$.
In the previous study by Ammi {\it et al.}, the relation between $\overline{n}$ and $V_I$ was obtained from
\begin{equation}
\label{eq:num}
\overline{n}(V_I)\sim n_0(1-\overline{e_R}^2)\left[\frac{V_I}{\zeta\sqrt{gd}}-1\right],
\end{equation}
where $n_0$ and $\zeta$ are the fitting parameters.
Our numerical results fit well with Eq. (\ref{eq:num}), where the values of the parameter pair are $(n_0,\zeta) \approx$ (43,\hspace{0.5mm}26) for the RP bed and $(n_0,\zeta)\approx (18,\hspace{0.5mm}28)$ for the FCC bed (Fig. \ref{num}).
The value of $\overline{n}$ for the case of the RP bed is more than twice that for the FCC bed.
This reflects the facts that the grains in the FCC bed experience a stronger geometrical constraint from the neighboring grains than those in the RP bed because of the higher volume fraction of the former, and $\overline{E_b}$ for the FCC bed is less than that for the RP bed in all pairs of $V_I$ and $\theta_I$ (Fig \ref{Energy}).

\begin{figure}[t]
\begin{center}
\includegraphics[scale=0.14]{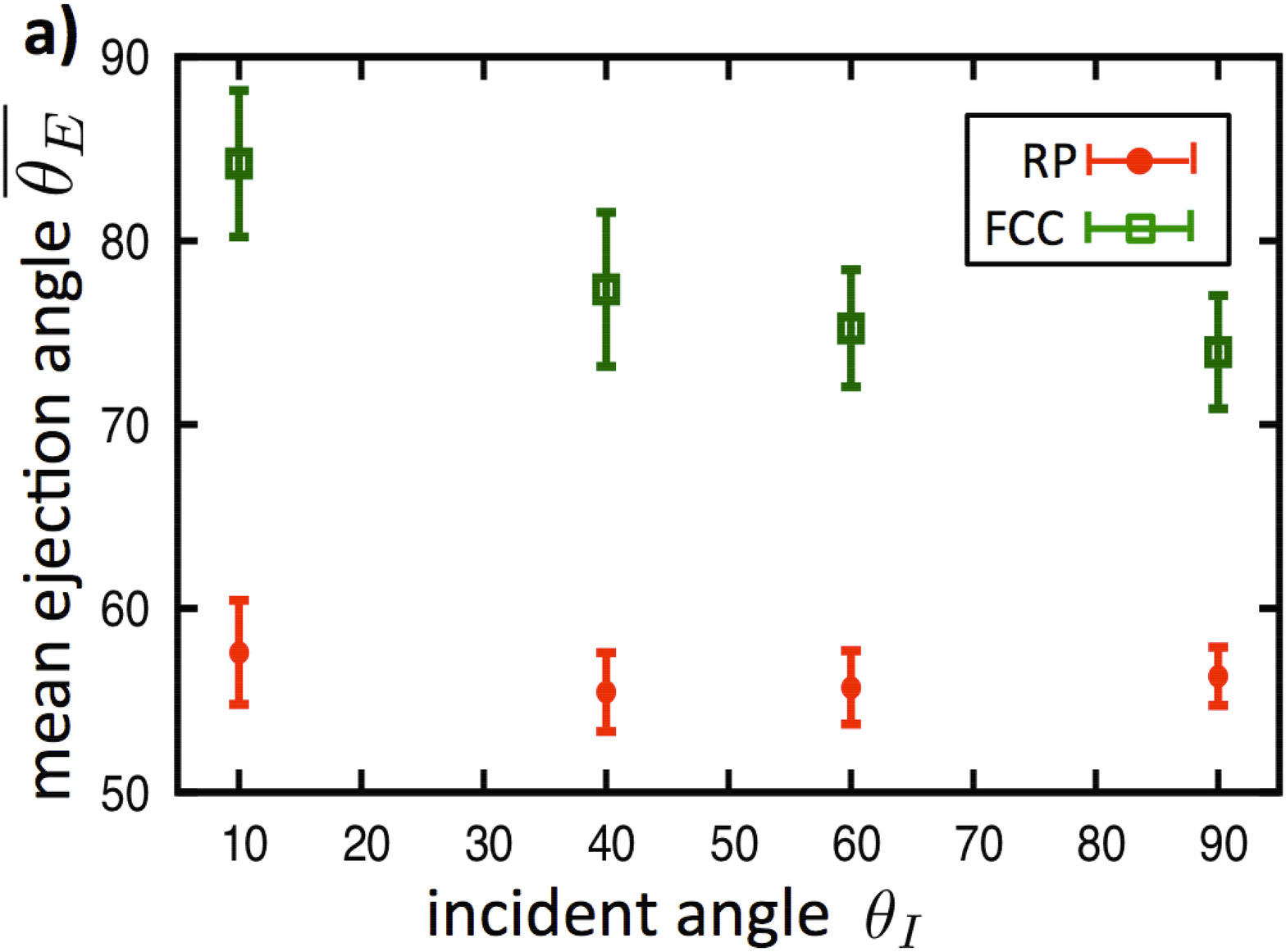}
\includegraphics[scale=0.14]{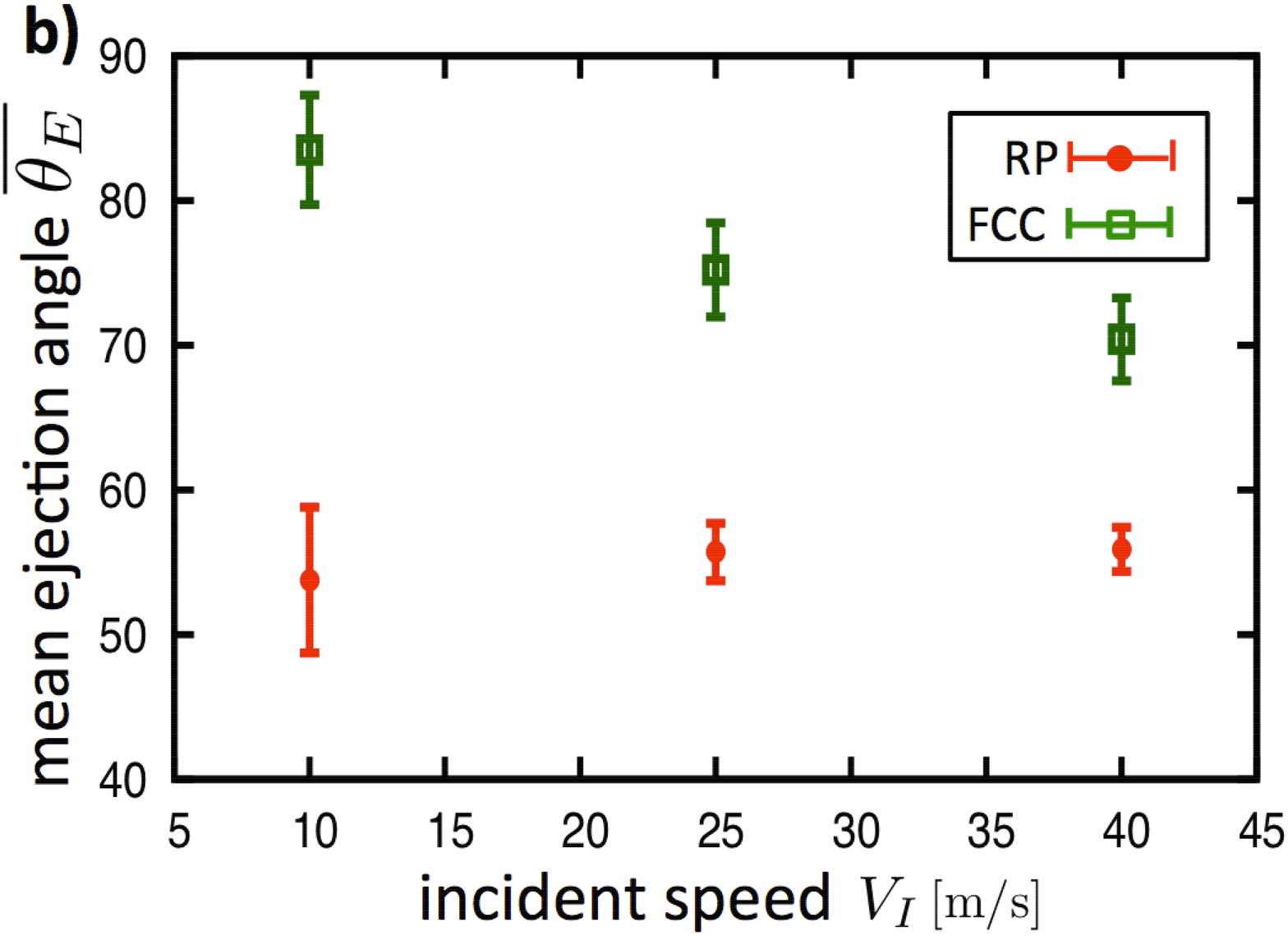}
\end{center}
\caption{
Mean ejection angle $\overline{\theta_E}$ for various (a) incident angles $\theta_I$ ($V_I=25.0{\rm m/s}$: fixed) and (b) incident speeds $V_I$ ($\theta_I=60^\circ$: fixed) in RP (filled circles) and FCC (open squares) beds.
The error bars indicate standard deviations.}
\label{AngMean}
\end{figure}
\subsection{Ejection Angle}
Figure \ref{AngMean} shows the ensemble averages of the mean ejection angle for each splash $\overline{\theta_E}$ for various values of $\theta_I$ (Fig. \ref{AngMean}(a)) and $V_I$ (Fig. \ref{AngMean}(b)).
According to the previously reported RP bed experiment\cite{Ammi}, $\overline{\theta_E}$ remains constant as $\theta_I$ and $V_I$ are varied.
Figure \ref{AngMean}(a) shows the $\theta_I$ dependence of $\overline{\theta_E}$ for $V_I=25$ m/s.
In this figure, our $\overline{\theta_E}$ for the RP bed remains almost constant and independent of $\theta_I$, which is consistent with the previous experiment\cite{Ammi}.
On the other hand, the $\overline{\theta_E}$ for the FCC bed clearly varies with $\theta_I$, especially at low $\theta_I$ (Fig. \ref{AngMean}(a)).
For the FCC bed, $\overline{E_b}$ become small at low $\theta_I$ (Fig. \ref{Energy}). 
This means that the FCC bed obtains insufficient energy to break the geometric constraint caused by the presence of the neighboring grains; hence, the ejection directions are strongly limited to high angles. 
However, the result for the FCC bed has not been confirmed experimentally.
Figure \ref{AngMean}(b) shows the $V_I$ dependence of $\overline{\theta_E}$ for fixed $\theta_I=60^\circ$.
For the RP bed, only a weak dependence is observed at low $V_I$, although this has not been confirmed experimentally \cite{Ammi}.
On the other hand, $\overline{\theta_E}$ exhibits an obvious dependence on $V_I$ for the FCC bed. 
That is, $\overline{\theta_E}$ decreases as $V_I$ increases.
This is attributed to the magnitude of $\overline{E_b}$, as discussed above.

\begin{figure}[tb]
\begin{center}
\includegraphics[scale=0.15]{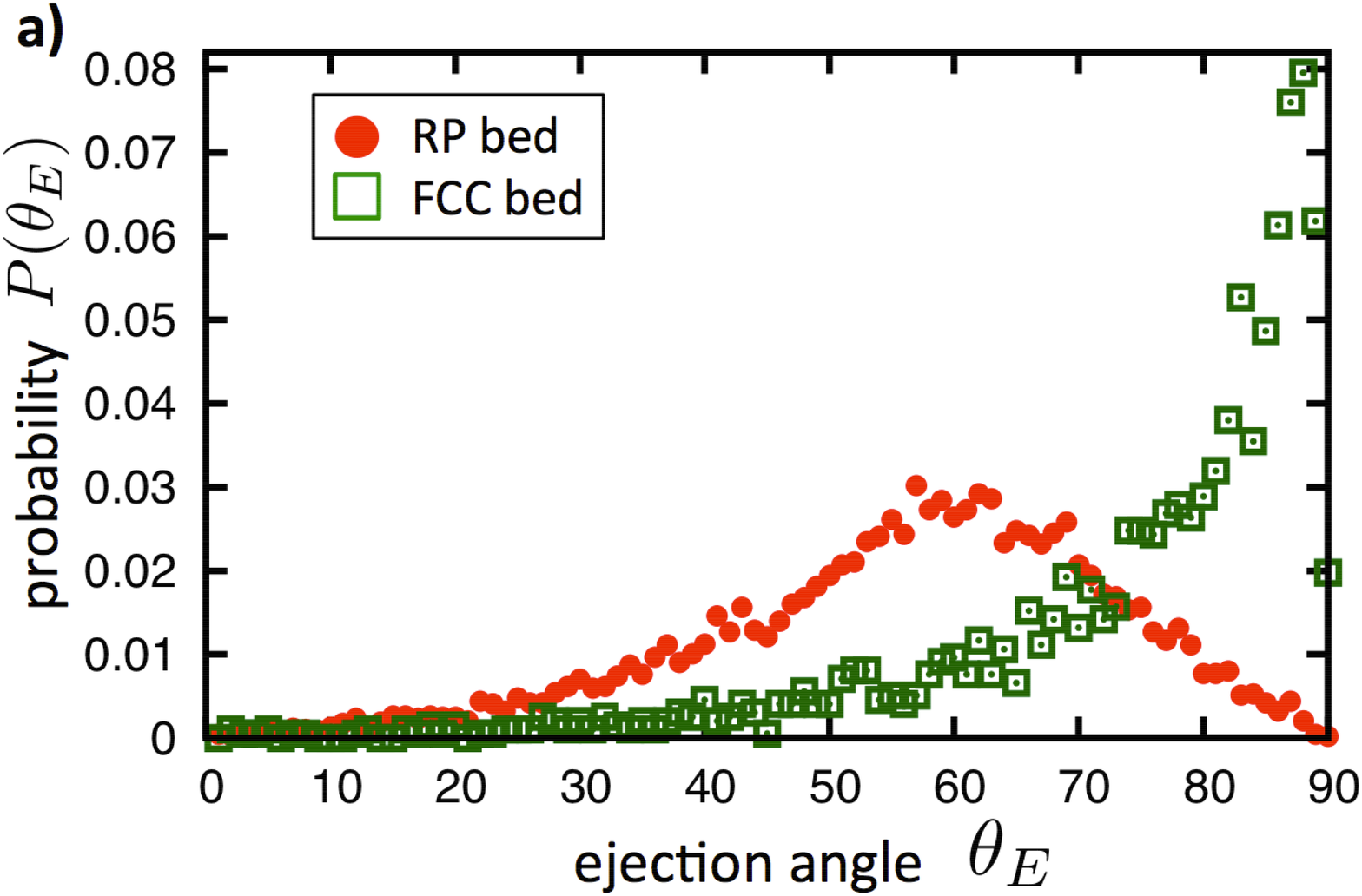}
\includegraphics[scale=0.175]{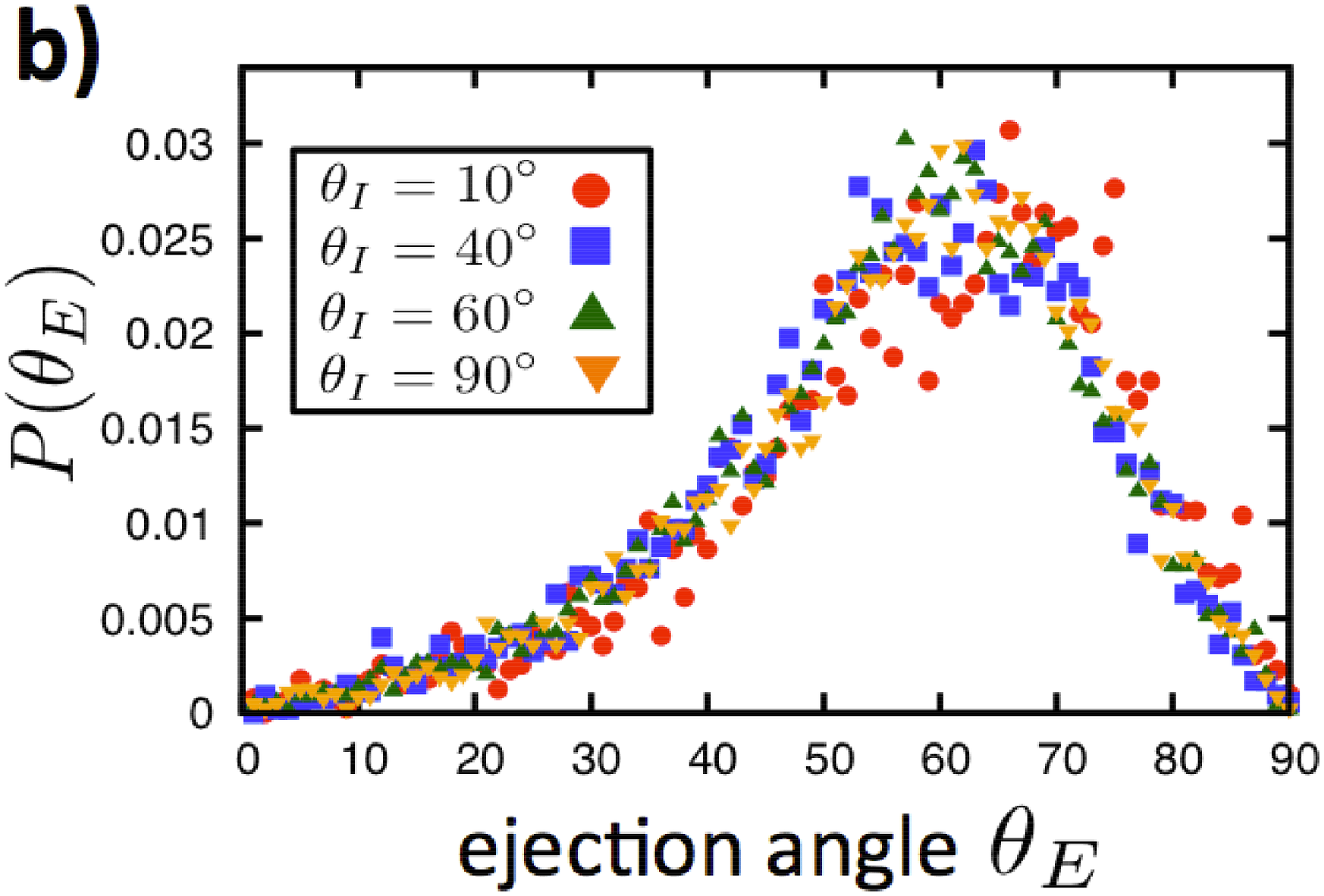}
\includegraphics[scale=0.175]{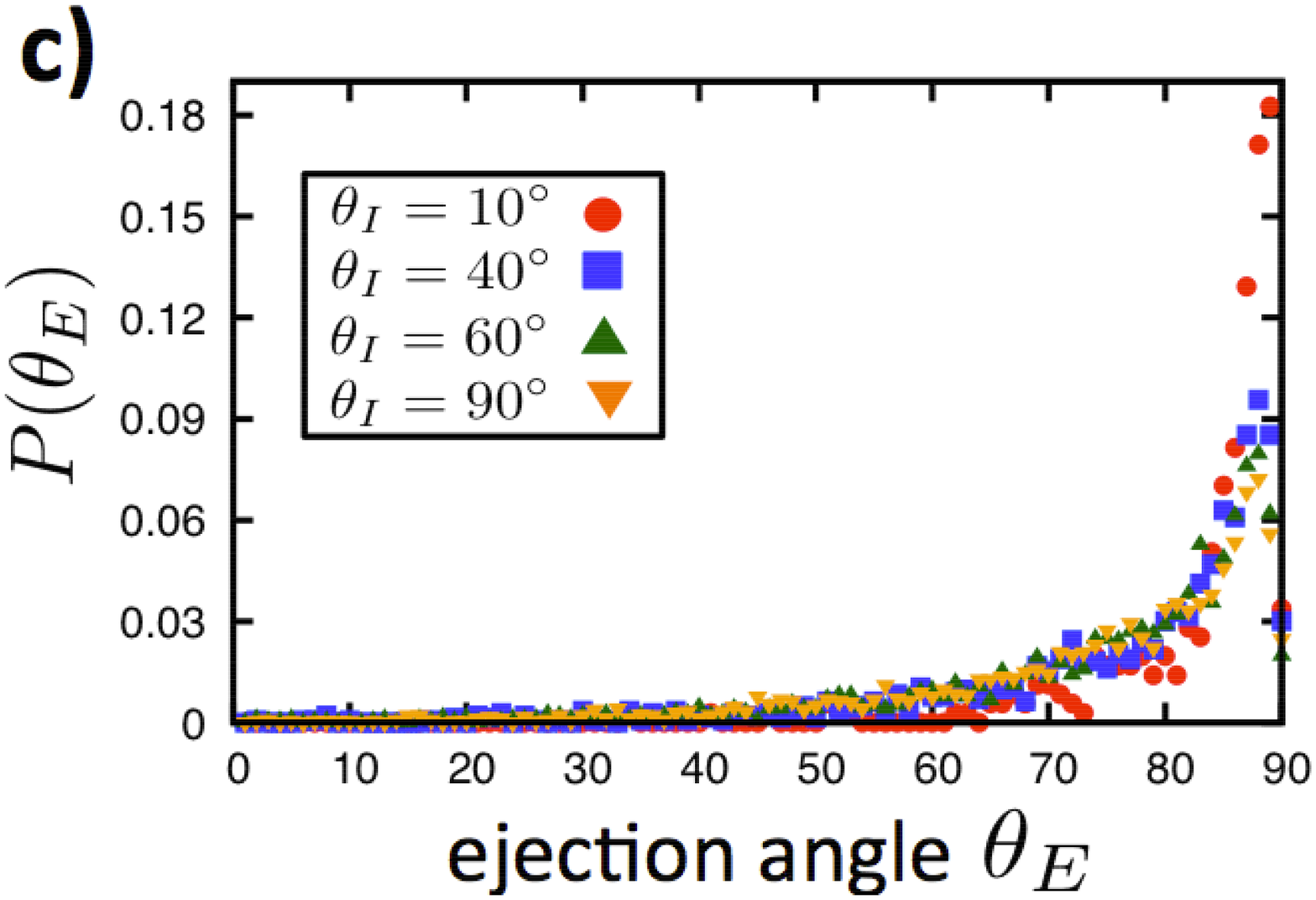}
\end{center}
\caption{
(a) Ejection angle distributions $P(\theta_E)$ for RP (filled circles) and FCC (open squares) beds.
The incident angle $\theta_I$ and incident speed $V_I$ are fixed ($\theta_I=60^\circ$ and $V_I=25.0{\rm m/s}$, respectively).
$P(\theta_E)$ for (b) RP and (c) FCC beds for various values of $\theta_I$ ($V_I=25.0$ m/s: fixed).
}
\label{f2}
\end{figure}
The ejection angle distributions $P(\theta_E)$ are shown in Fig. \ref{f2}.
$P(\theta_E)$ for the RP bed obviously differs from that obtained for the FCC bed.
The majority of grains ejected from the FCC bed have greater $\theta_E$ than those ejected from the RP bed (Fig. \ref{f2}(a)).
On the other hand, $P(\theta_E)$ for the RP bed is independent of $\theta_I$, and the shapes and locations of the peaks around $60^\circ$ exhibit good agreement with the findings of a previous numerical experiment using binary grains\cite{Mao} (Fig. \ref{f2}(b)).

\begin{figure}[t]
\includegraphics[scale=0.195]{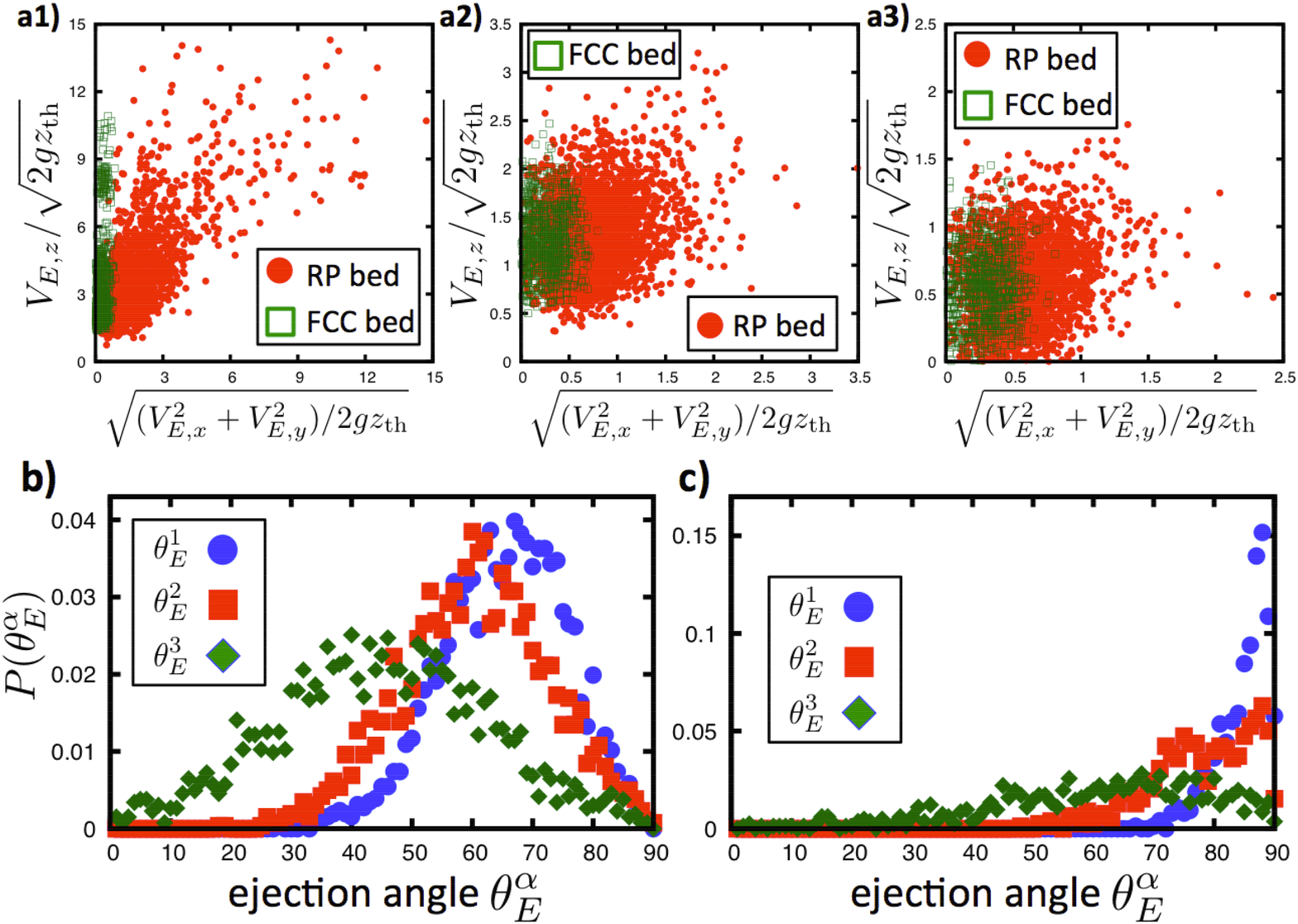}
\caption{
(a) Scatter plot of the grains in the $G_1, G_2$, and $G_3$ groups on the $V_{E,x,y}-V_{E,z}$ plane.
The filled circles and open squares represent grains ejected from the RP and FCC beds, respectively.
Ejection angle distributions $P(\theta_E^\alpha$) for (b) RP and (c) FCC beds.
$\theta_E^\alpha$ is the ejection angle of $G_\alpha$ ($\alpha = \{1, 2, 3\}$, $V_I=25.0\,{\rm m/s}$, and $\theta_I=90^\circ$).
}
\label{ProbOrder}
\end{figure}

To investigate each splash process in greater detail, we classify the ejected grains into three groups on the basis of their ejection timing.
The first group $G_1$ consists of grains that were ejected in the period between the moment of impact and the first third of the total ejection period of each splash process.
The ejection angles of the particles in this group are labeled $\theta_E^1$.
Similarly, the ejection angles of the grains in groups $G_2$ and $G_3$, which were ejected within the intermediate period and the last third of each splash process, respectively, are labeled $\theta_E^2$ and $\theta_E^3$, respectively.
Figure \ref{ProbOrder} shows scatter plots for grains belonging to the $G_1, G_2$, and $G_3$ groups on the $V_{E,x,y}-V_{E,z}$ plane, where $V_{E,x,y}$ indicates the projection of $\mbox{\boldmath$V$}_E$ onto the bed surface.
The ejection angle is defined as the angle between the horizontal axis and the line connecting the origin and each point in Fig. \ref{ProbOrder}(a), which indicates that the magnitude of $\mbox{\boldmath$V$}_E$ varies depending on the ejection timing.
The distributions of $\theta_E^1, \theta_E^2$, and $\theta_E^3$ ($P(\theta_E^1)$, $P(\theta_E^2)$, and $P(\theta_E^3)$, respectively) for $V_I=25.0{\rm m/s}$ and $\theta_I=90^\circ$ are also shown for both bed types (Fig. \ref{ProbOrder}(b) and (c)).

Since the peaks of $P(\theta_E^1)$ and $P(\theta_E^2)$ obtained for the RP bed and those for the FCC bed are at greater angles, these grains seem to be affected by their neighboring grains.
This is particularly true in the FCC case (Fig. \ref{ProbOrder}(c)), where the grain movements are obviously restricted to the higher angles: both $P(\theta_E^1)$ and $P(\theta_E^2)$ have peaks around $90^\circ$, but the peak of $P(\theta_E^1)$ is higher than that of $P(\theta_E^2)$.
The profiles of $P(\theta_E^3)$ for both bed types are different than those of $P(\theta_E^1)$ and $P(\theta_E^2)$; the $P(\theta_E^3)$ peaks are clearly located within a lower range of angles compared to those of $P(\theta_E^1)$ and $P(\theta_E^2)$.
As supported by the discussion of the $V_{E,z}$ results in the next section, these results for the FCC bed suggest that the grain ejection direction is more strongly restricted by geometrical constraints compared to the RP bed.

\begin{figure}[tb]
\includegraphics[scale=0.14]{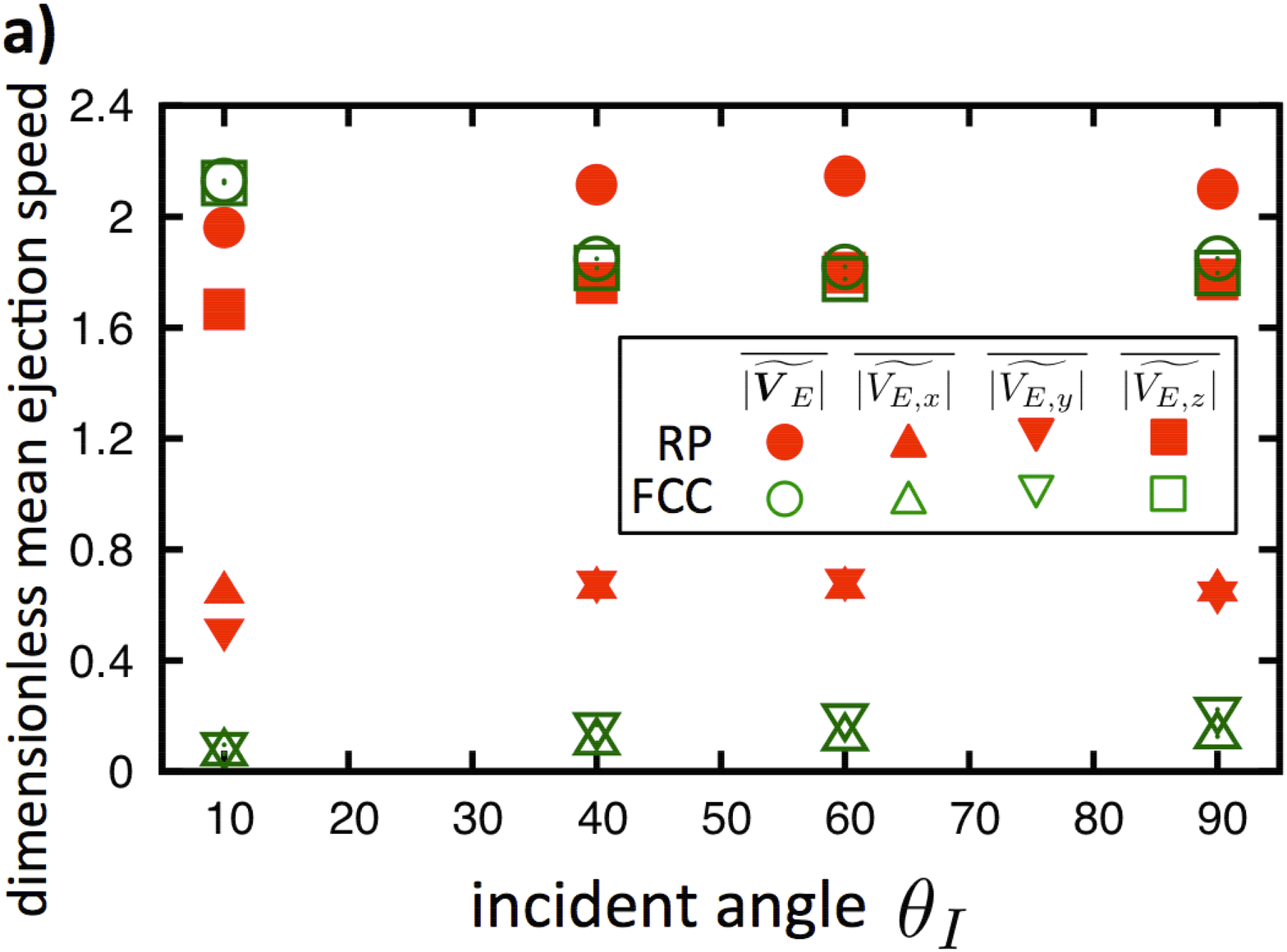}
\includegraphics[scale=0.14]{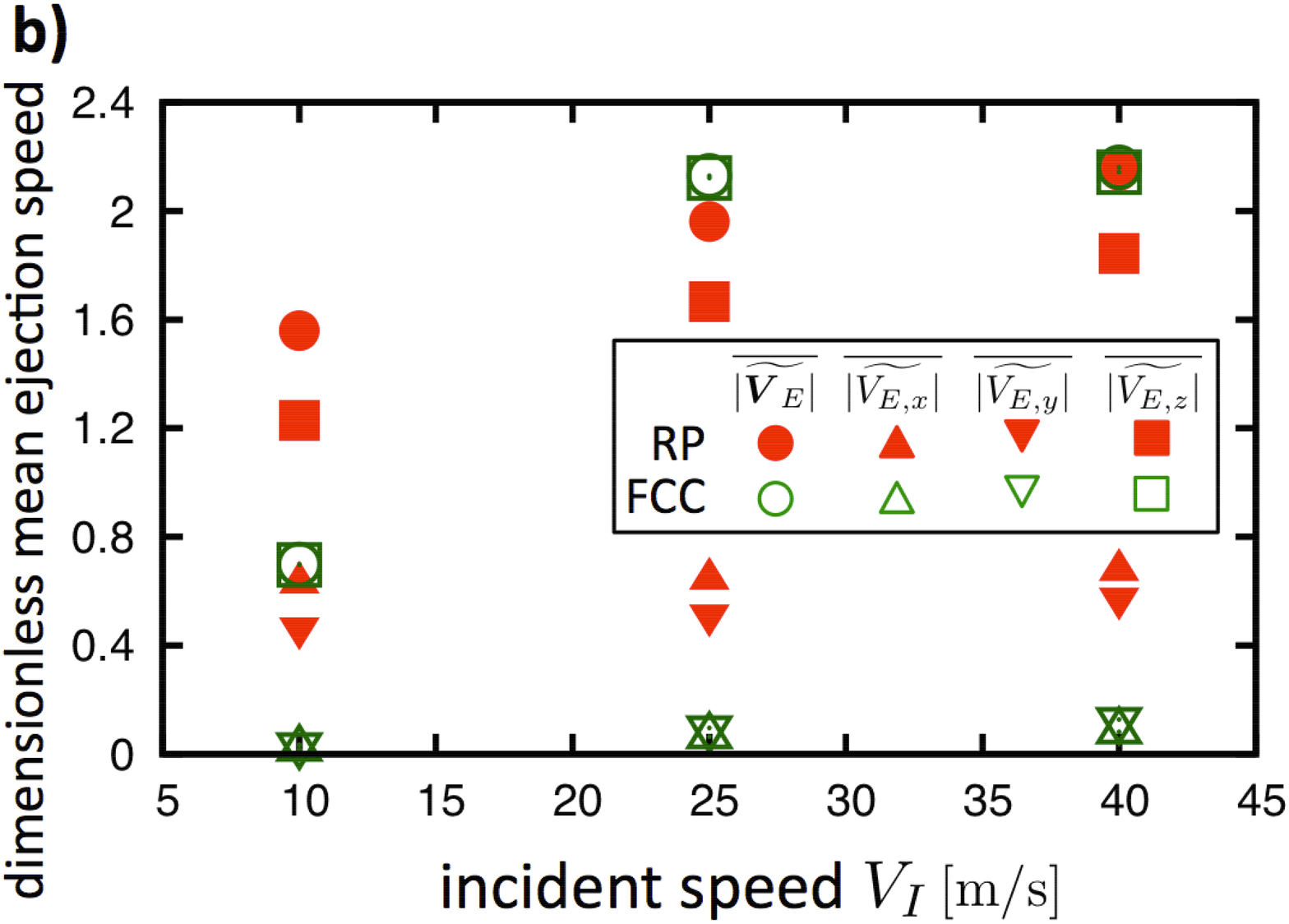}
\caption{
Normalized mean ejection speed $\overline{|\widetilde{\mbox{\boldmath$V$}_E}|}$, $\overline{|\widetilde{V_{E,x}}|}$, $\overline{|\widetilde{V_{E,y}}|}$, and $\overline{|\widetilde{V_{E,z}}|}$, for (a) various incident angles $\theta_I$ ($V_I=25.0{\rm m/s}$) and (b) incident speeds $V_I$ ($\theta_I=10^\circ$) in RP (filled symbols) and FCC (open symbols) beds, where $\widetilde{\mbox{\boldmath$V$}_E}=(\widetilde{V_{E,x}},\widetilde{V_{E,y}},\widetilde{V_{E,z}})=$$(V_{E,x}/\sqrt{2gz_{\rm th}}$$,V_{E,y}/\sqrt{2gz_{\rm th}},$$V_{E,z}/\sqrt{2gz_{\rm th}})$.}
\label{MeanVel}
\end{figure}
\subsection{Ejection Velocity}
Figure \ref{MeanVel} shows the ensamble averages of the mean ejection speed for each splash $\overline{|\widetilde{\mbox{\boldmath$V$}_E}|}$ and its components $\overline{|\widetilde{V_{E,\beta}}|}$ $(\beta\in\{x,y,z\})$ for various values of $\theta_I$ (Fig. \ref{MeanVel}(a)) and $V_I$ (Fig. \ref{MeanVel}(b)), where $\widetilde{\mbox{\boldmath$V$}_E}$$=$$(\widetilde{V_{E,x}},$$\widetilde{V_{E,y}},$$\widetilde{V_{E,z}})$$=(V_{E,x}/\sqrt{2gz_{\rm th}},$$V_{E,y}/\sqrt{2gz_{\rm th}},$$V_{E,z}/\sqrt{2gz_{\rm th}})$.
For all pairs of $\theta_I$ and $V_I$, the greater part of $\overline{|\widetilde{\mbox{\boldmath$V$}_E}|}$ is $\overline{|\widetilde{V_{E,z}}|}$.
Figure \ref{MeanVel}(a) shows the $\theta_I$ dependence of $\overline{|\widetilde{\mbox{\boldmath$V$}_E}|}$ and $\overline{|\widetilde{V_{E,\beta}}|}$ for $V_I=25.0$ m/s.
Although there is a slight fluctuation within the low-incident-angle region, $\overline{|\widetilde{\mbox{\boldmath$V$}_E}|}$ remains almost constant as $\theta_I$ is varied for both bed structures.
In the RP bed, there is a small gap between $\overline{|\widetilde{V_{E,x}}|}$ and $\overline{|\widetilde{V_{E,y}}|}$ for $\theta_I=10^\circ$.
In contrast, $\overline{|\widetilde{V_{E,x}}|}$ and $\overline{|\widetilde{V_{E,y}}|}$ are the almost same for $\theta_I\ge40^\circ$.
Figure \ref{MeanVel}(b) shows the $V_I$ dependency of $\overline{|\widetilde{\mbox{\boldmath$V$}_E}|}$ and $\overline{|\widetilde{V_{E,\beta}}|}$ for $\theta_I=10^\circ$.
In this figure, the mean ejection speed increases as $V_I$ increases.
These $\theta_I$ and $V_I$ dependencies are consistent with a previous study\cite{Mao}.

\begin{figure}[t]
\begin{center}
\includegraphics[scale=0.18]{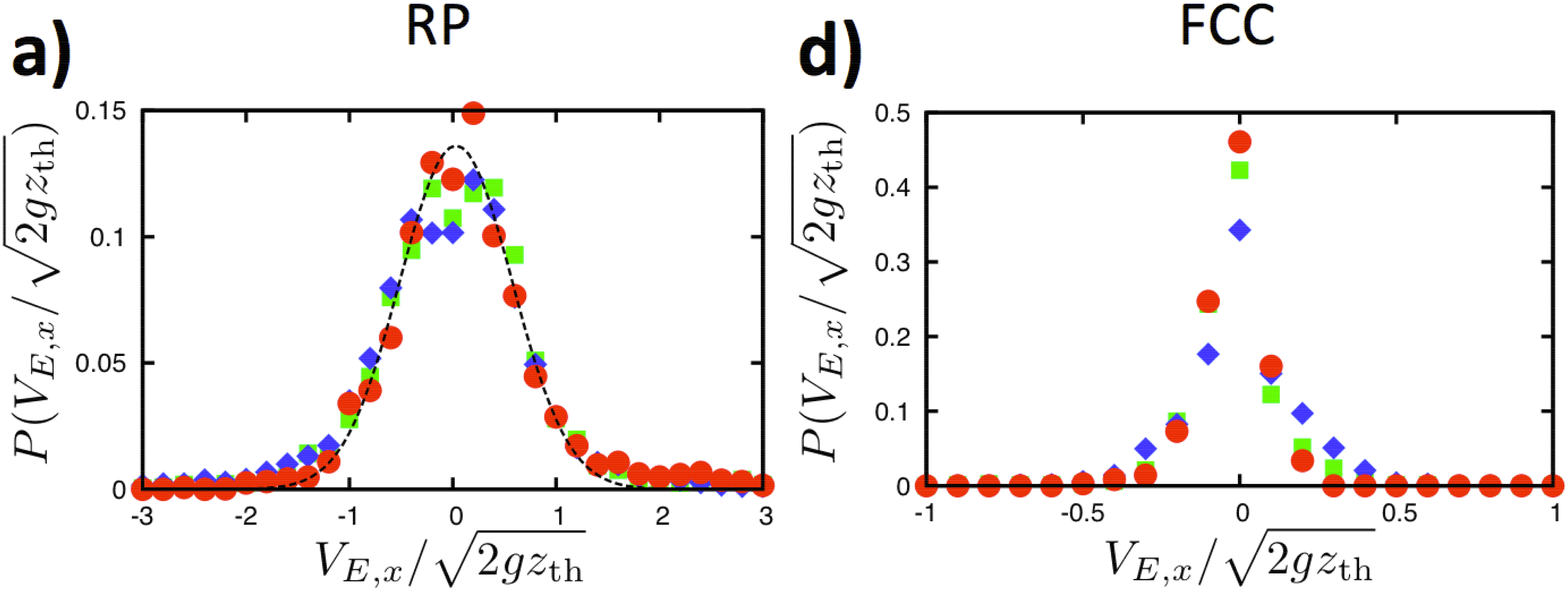}
\includegraphics[scale=0.18]{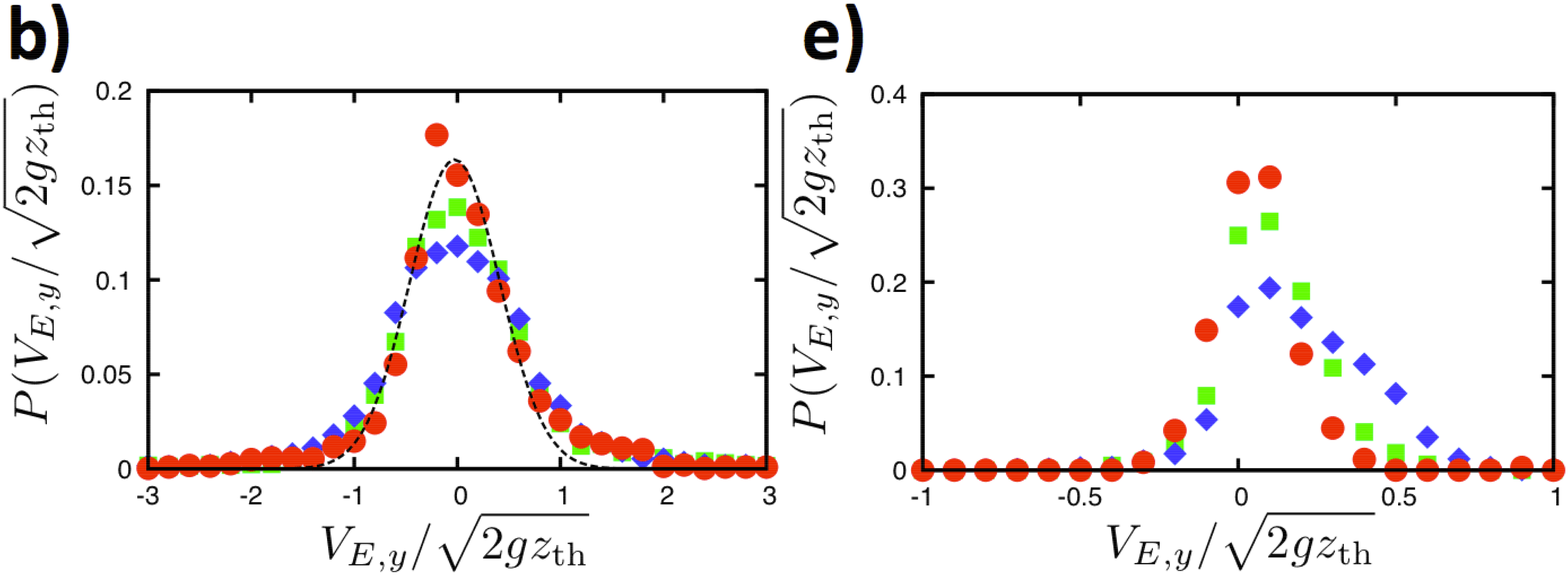}
\includegraphics[scale=0.18]{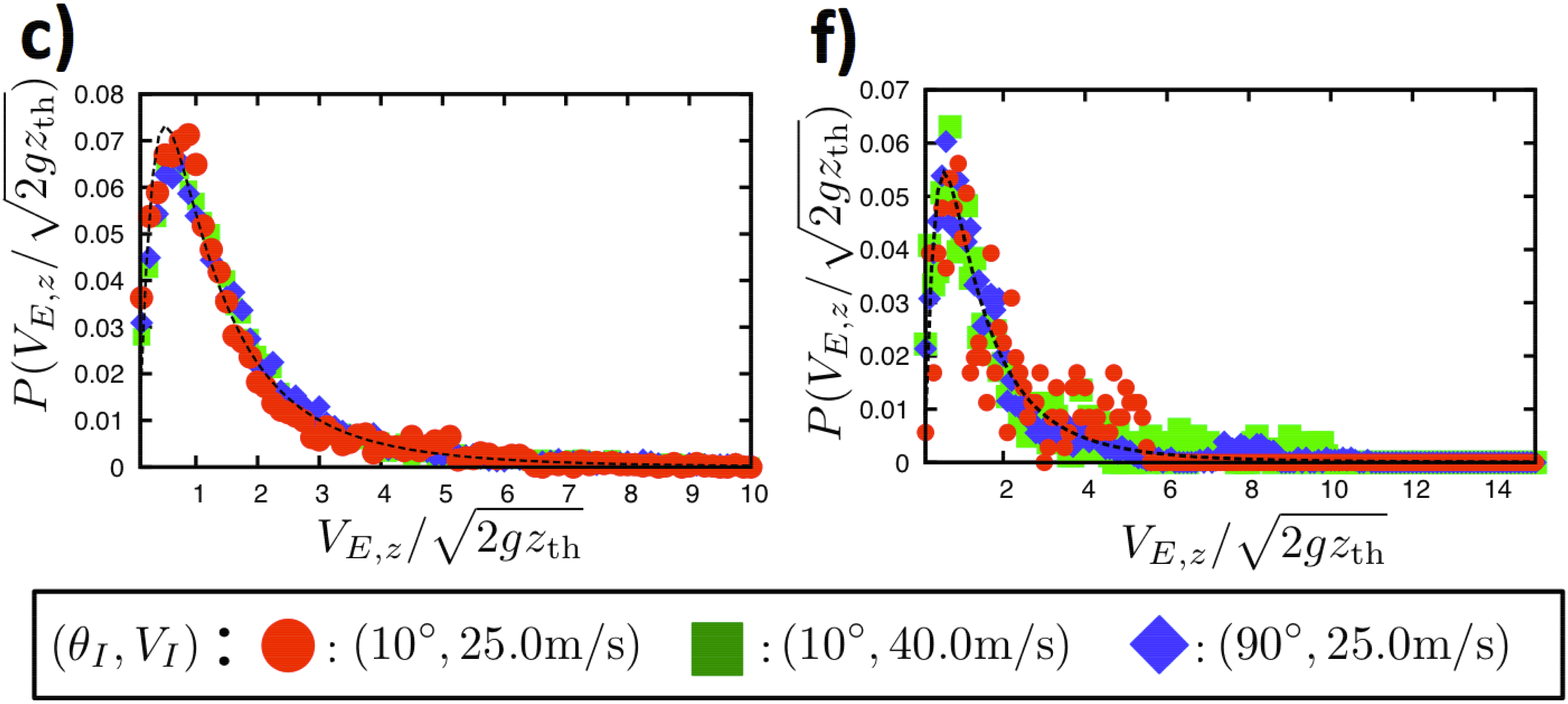}
\caption{
Ejection velocity distributions for (a) $V_x$, (b) $V_y$, and (c) $V_z$ obtained for the RP bed and those for (d) $V_x$, (e) $V_y$, and (f) $V_z$ obtained for the FCC bed for various incident angles $\theta_I$ and speeds $V_I$.
All values are normalized by $\sqrt{2gz_{\rm th}}$.
The dashed lines represent the best fit for each distribution for $\theta_I=90^\circ$ and $V_I=25.0$ m/s (filled circles).
}
\label{Vel}
\end{center}
\end{figure}

We next investigate the distributions of each component of $\widetilde{\mbox{\boldmath$V$}_E}$, $P(\widetilde{V_{E,x}})$, $P(\widetilde{V_{E,y}})$ and $P(\widetilde{V_{E,z}})$, for the different bed structures in Fig. \ref{Vel}.
For the RP bed, both $P(\widetilde{V_{E,x}})$ and $P(\widetilde{V_{E,y}})$ have Gaussian distributions (Fig. \ref{Vel} (a) and (b)), whereas $P(\widetilde{V_{E,z}})$ has a log-normal appearance (Fig. \ref{Vel} (c)).
These results are consistent with the findings of previous experimental studies\cite{Ammi,Rioul}.
Note that these forms are independent of both $\theta_I$ and $V_I$ (Fig. \ref{Vel}(a), (b) and (c)).
For the FCC bed, $P(\widetilde{V_{E,z}})$ appears to be similar to that obtained for the RP bed (Fig. \ref{Vel} (f)), but both $P(\widetilde{V_{E,x}})$ and $P(\widetilde{V_{E,y}})$ are more concentrated around 0 m/s than those for the RP bed (Fig. \ref{Vel} (d) and (e)).
Regarding the difference between the $P(\widetilde{V_{E,z}})$ for the RP and FCC beds, the latter has a bump within the large $V_{E,z}$ range (see also Fig. \ref{ProbOrder}(a1)).

\begin{figure}[t]
\begin{center}
\includegraphics[scale=0.18]{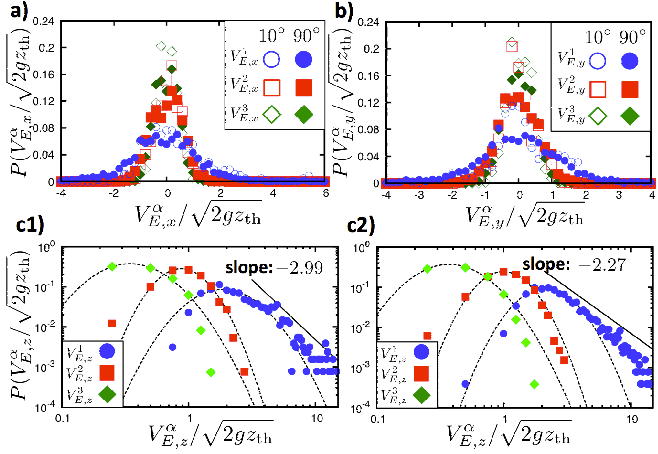}
\end{center}
\caption{
Normalized horizontal ejection speed distributions (a) $P(V_{E,x}^\alpha/\sqrt{2gz_{\rm th}})$ and (b) $P(V_{E,y}^\alpha/\sqrt{2gz_{\rm th}})$ for $\theta_I=10^\circ$ (open symbols) and $\theta_I=90^\circ$ (closed symbols), and vertical ejection speed distributions $P(V_{E,z}^\alpha/\sqrt{2gz_{\rm th}})$ for (c1) $\theta_I=10^\circ$ and (c2) $\theta_I=90^\circ$ for the RP bed ($V_I=25.0$ m/s: fixed).
$\mbox{\boldmath$V$}_{E}^\alpha$ is the ejection velocity of $G_\alpha$ ($\alpha\in\{1,2,3\}$).
The dashed lines represent the fits obtained for a log-normal distribution.
}
\label{DivVel}
\end{figure}

We also define the timing-dependent ejection velocities in conformity to the groups $G_1$, $G_2$, and $G_3$, as $\mbox{\boldmath$V$}_E^1$, $\mbox{\boldmath$V$}_E^2$, and $\mbox{\boldmath$V$}_E^3$, respectively.
Figure \ref{DivVel} shows the vertical ejection speed distributions $P(\widetilde{V_{E,z}^1})$,  $P(\widetilde{V_{E,z}^2})$, and  $P(\widetilde{V_{E,z}^3})$ obtained for $\theta_I=10^\circ$ and $\theta_I=90^\circ$ for $V_I=25.0$ m/s in the RP bed. 
For all $\alpha\in\{1,2,3\}$, $\widetilde{V_{E,x}^\alpha}$ and $\widetilde{V_{E,y}^\alpha}$ have Gaussian-like form, but their forms are different and depend on the ejection timing; $P(\widetilde{V_{E,x}^1})$ and $P(\widetilde{V_{E,y}^1})$ have large variances, and the others have small variances (Fig. \ref{DivVel}(a) and (b)).
$P(\widetilde{V_{E,z}^2})$ and $P(\widetilde{V_{E,z}^3})$ fit well with the log-normal distributions, but the higher-ejection-speed region of $P(\widetilde{V_{E,z}^1})$ seems to have a power-law form (Fig. \ref{DivVel}(c)). 
That is, the distributions change from a power-law form to a log-normal form as the ejection speed is decreases (or with increasing elapsed time since impact).
As this power-law region is only a small fraction of the total vertical ejection speed distribution, the overall distribution $P(\widetilde{V_{E,z}})$ throughout each splash process is fit well with a log-normal distribution.
This distribution deformation becomes clear with increasing incident angle.
Further, these types of distribution transformations have been reported in various fields.
For example, fragment experiments have confirmed that the fragment size distribution of glass qualitatively changes from a log-normal distribution to power-law form in accordance with the incident energy\cite{Matsushita,Katsuragi2}. 
Specifically, log-normal and power-law distributions are exhibited at lower and higher energies, respectively.
Therefore, our results may be related to these findings.

\subsection{Ejection Energy}
\begin{figure}[tb]
\begin{center}
\includegraphics[scale=0.18]{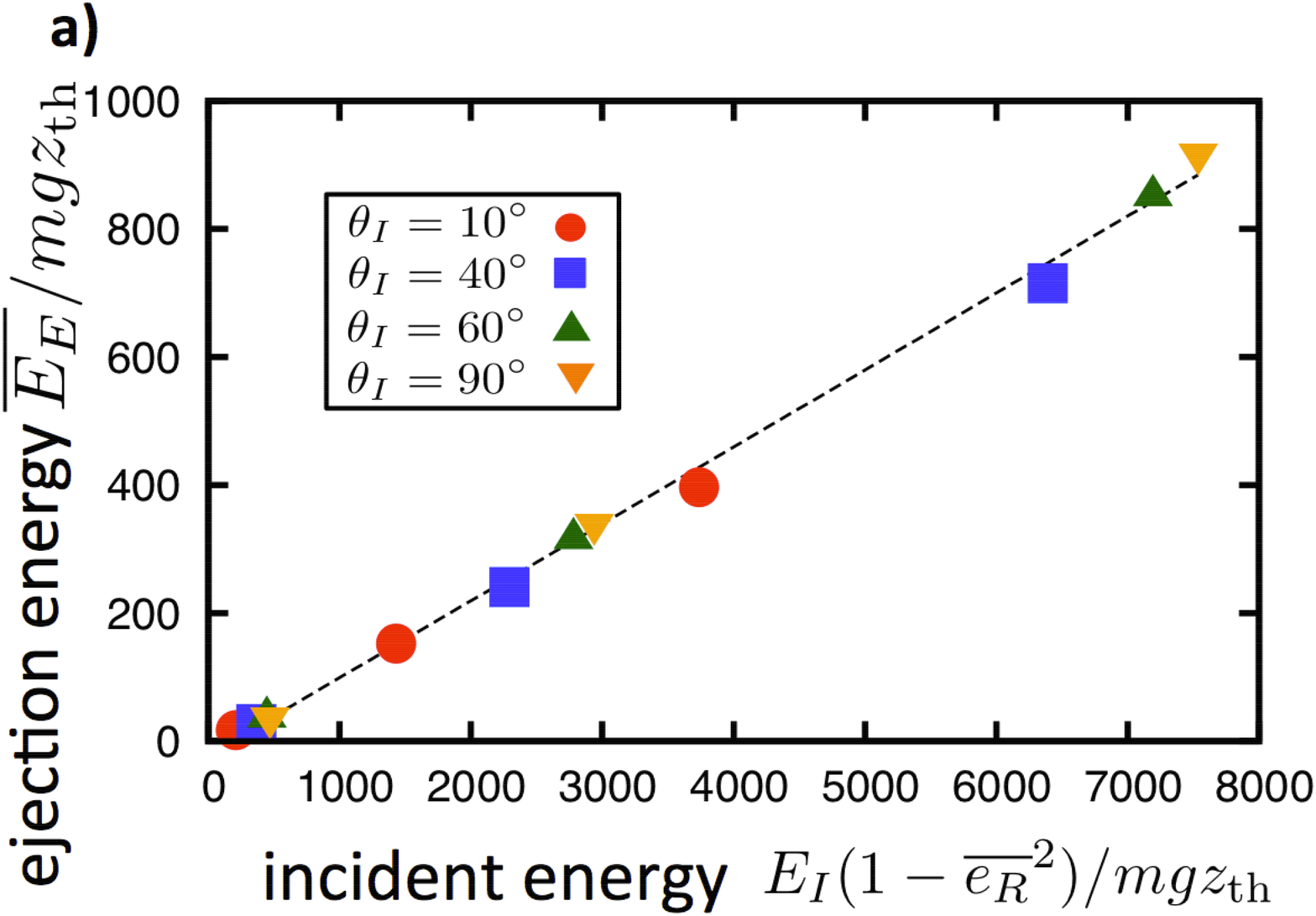}
\includegraphics[scale=0.115]{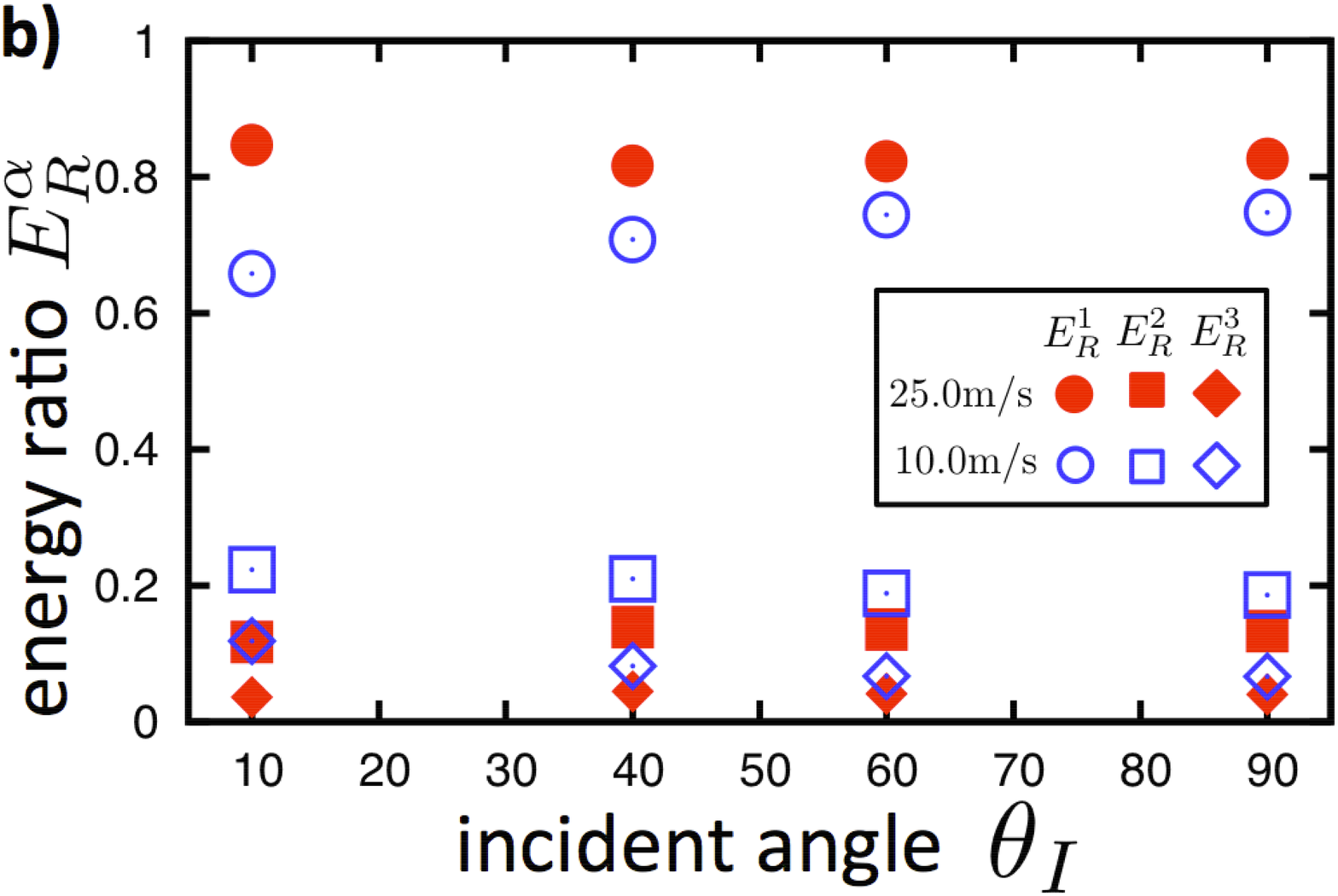}
\includegraphics[scale=0.115]{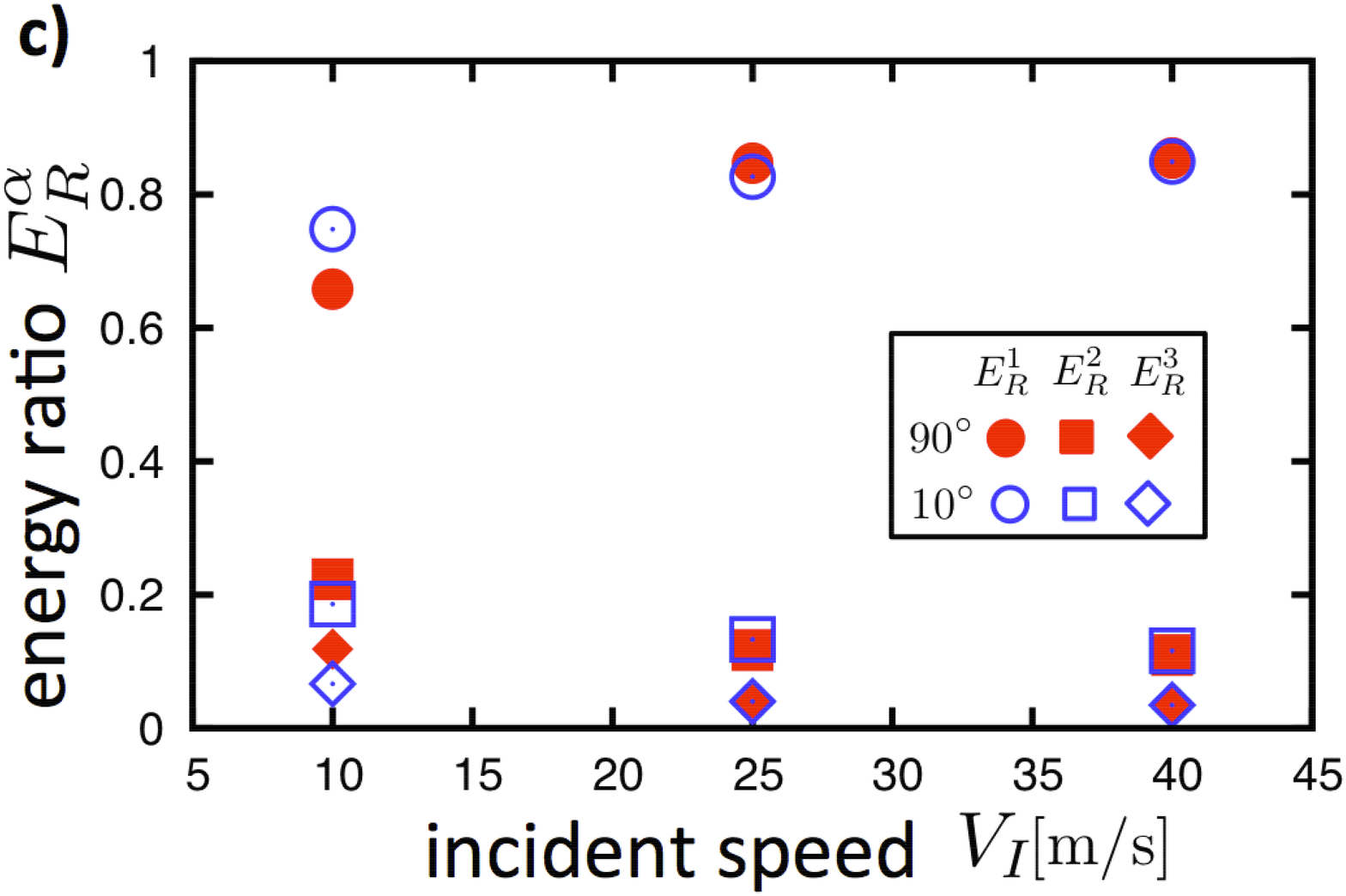}
\end{center}
\caption{
(a)The relation between the total kinetic energy of the splashed grains and the fraction of energy transferred into the granular bed $E_I(1-\overline{e_R}^2)$.
All values are normalized by $mgz_{\rm th}$.
Energy ratio $E_R^\alpha=\overline{E_E^\alpha}/\left(\overline{E_E^1}+\overline{E_E^2}+\overline{E_E^3}\right)$ for (b) various incident angles $\theta_I$ ($V_I=10.0{\rm m/s}$: open symbols and $V_I=25.0{\rm m/s}$: closed symbols) and (c) incident speeds $V_I$ ($\theta_I=10^\circ$: open symbols and $\theta_I=90^\circ$: filled symbols) $(\alpha=\{1,2,3\})$.
All points are obtained for the RP bed.
}
\label{EneBalance}
\end{figure}

We show the energy balances in Fig. \ref{EneBalance}.
The energy balance between the incident energy $\overline{E_b}$ and the total kinetic energy of the ejected grains $\overline{E_E}=m\overline{n}\overline{|\mbox{\boldmath$V$}_E|}^2/2$ is shown in Fig. \ref{EneBalance}(a).
As noted from previous experiments\cite{Ammi}, the relation between $\overline{E_b}$ and $\overline{E_E}$ is $\overline{E_E}\approx r\overline{E_b}$, where $r$ is a constant parameter ($r\approx0.12$ in our result).
Because the rotational motion of a grain is not considered in this study, that is, the obtained kinetic energy reflects only translational motion, $r$ in our study may be greater than $r$ in the experiment of Ammi {\it et al} ($r\approx0.04$)\cite{Ammi}.
Previously, it was found that $r$ depends on the restitution coefficient in a binary collision\cite{Crassous}.

Figure \ref{EneBalance}(b) and (c) show the energy ratio $E_R^\alpha=\overline{E_E^\alpha}/\left(\overline{E_E^1}+\overline{E_E^2}+\overline{E_E^3}\right)$ for the RP bed, where $\overline{E_E^\alpha}=m\overline{n_E^\alpha}\overline{V^\alpha_E}^2/2$ is the total kinetic energy of ejected grains belonging to $G_\alpha$, and $\overline{n^\alpha}\approx\overline{n}/3$ is mean number of ejected grains per impact for $G_\alpha$. 
Figure \ref{EneBalance}(b) shows the $\theta_I$ dependence of $E_R^\alpha$ for $V_I=10.0$ m/s and $V_I=25.0$ m/s, and Fig. \ref{EneBalance}(c) shows the $V_I$ dependence of $E_R^\alpha$ for $\theta_I=10^\circ$ and $\theta_I=90^\circ$.
In these figures, $E_R^\alpha$ is almost independent of $\theta_I$; in particular, for larger values of $V_I$, the values of $E_R^\alpha$ for $\theta_I=10^\circ$ and $\theta_I=90^\circ$ are mostly coincident, and more than 80\% of the total ejection energy is used for $G_1$ grains.

\section{Summary}
We performed 3D splash process simulations using the DEM for two kinds of granular bed structures: a randomly structured bed and an FCC-structured bed. 
It was found that the mean number of ejected grains for each collision was related to the injection energy.
After renormalization by the energy transferred from the incident grain to the granular bed, a good linear fit was obtained between the mean number of ejected grains and the incident speed, with the RP bed ejecting twice as many grains as the FCC bed.
Moreover, the ejection angle distributions obtained from the RP and FCC beds were shown to be clearly different.
The peak of the ejection angle distribution for the RP bed was approximately $60^\circ$; on the other hand, the distribution obtained for the FCC bed distinctively shifted to greater ejection angles, with a peak of over $80^\circ$.
This difference is assumed to originate from the geometrical constraints. 
In other words, the grain movement direction is strongly affected by the surrounding grains in the FCC bed.
Furthermore, the ejection velocity distributions for the RP bed exhibited qualitatively good agreement with the results of previous experiments\cite{Ammi}.
On the other hand, coupled with the ejection angle results, the distributions obtained for the FCC bed indicate that the vertical movement of the ejected grains is dominant and that movement in the horizontal direction is significantly smaller than that for the RP bed.

In addition, the ejected-grain characteristics, i.e., the ejection angle and speed, evidently depend on the ejection timing after the initial grain impact.
For the ejection angle, the difference between the ejected grain angles at the beginning and end of each splash is apparent.
Regarding the vertical ejection speed, the ejection timing determines the distribution, and this distribution changes from a power-law form to a log-normal form according to the ejection timing.
Furthermore, the splashed grains at the beginning of each splash gain retains around 80\% of the total kinetic energy of the ejected grains.
These results are assumed to be related to the propagation of the impact energy, both along and beneath the surface of the granular bed.

\begin{acknowledgments}
The authors thank A. Awazu and H. Niiya for useful discussions.
This research is partially supported by the Platform Project for Supporting in Drug Discovery and Life Science Research (Platform for Dynamic Approaches to Living System) from Japan Agency for Medical Research and Development (AMED)
\end{acknowledgments}

\end{document}